\documentclass[prl,superscriptaddress,a4paper,twocolumn,showpacs]{revtex4}
\usepackage{amsmath,amsfonts,graphicx,color}
\usepackage{amssymb}
\usepackage{epsfig}

\renewcommand{\>}{\rangle}

\begin{document}

\title{Qubit-Initialisation and Readout with Finite Coherent Amplitudes in Cavity QED}
\author{Ricardo Kennedy}%
%
\affiliation{Optics Section, Blackett Laboratory, Imperial College London, London SW7 2BZ, United Kingdom}%
\affiliation{Institute for Mathematical Sciences, Imperial College London, London SW7 2BW, United Kingdom}%
\author{Leonhard Horstmeyer}%
\affiliation{Optics Section, Blackett Laboratory, Imperial College London, London SW7 2BZ, United Kingdom}%
\affiliation{Institute for Mathematical Sciences, Imperial College London, London SW7 2BW, United Kingdom}%

\author{Andrzej Dragan}%

\affiliation{Optics Section, Blackett Laboratory, Imperial College London, London SW7 2BZ, United Kingdom}%
\affiliation{Institute of Theoretical Physics, University of Warsaw, Ho\.{z}a 69, 00-681 Warsaw, Poland}%

\author{Terry Rudolph}%

\affiliation{Optics Section, Blackett Laboratory, Imperial College London, London SW7 2BZ, United Kingdom}%
\affiliation{Institute for Mathematical Sciences, Imperial College London, London SW7 2BW, United Kingdom}%


\begin{abstract}
We consider a unitary transfer of an arbitrary state of a two-level atomic qubit in a cavity to the finite amplitude coherent state cavity field. Such transfer can be used to either provide an effective readout measurement on the atom by a subsequent measurement on the light field or as a method for initializing a fixed atomic state - a so-called ``attractor state'', studied previously for the case of an infinitely strong cavity field. We show that with a suitable adjustment of the coherent amplitude and evolution time the qubit transfers all its information to the field, attaining a selected state of high purity irrespectively of the initial state.
\end{abstract}

\maketitle

\section{Introduction}

Atom-light interaction in the Jaynes-Cummings Model \cite{jaynescummings} has been studied
extensively over the last few decades. In particular the interaction of a light field initially in a coherent state with a single two-level atom in a cavity attracted much interest, due to counter-intuitive features like the collapse and revival of the atomic energy eigenstate oscillations. This effect has been extensively studied by Phoenix and Knight \cite{Phoenix} and later by Gea-Banacloche \cite{gea,attractor}, who introduced the so-called ``attractor states''. It is with this latter phenomenon that we will be concerned.

Recently, the possibility of attractor states when multiple atoms are in the cavity - a case of relevance to quantum computation - has been studied \cite{attractor2}. These studies were undertaken mainly under the assumption of an infinitely strong coherent state. In this paper we will return to the case of a single qubit. The idealized process which we are interested in is therefore an evolution of an arbitrary qubit state $|\psi\>$, coupled by the Jaynes-Cummings Hamiltonian to a single-mode field initially in a fixed coherent state $|\alpha\>$, such that the atom is left in a final, known, fixed state $|0\>$ while the field is left in some new state $|\alpha'(\psi)\>$. That is, ideally we implement the transformation:
\[
|\psi\>|\alpha\>\rightarrow |0\>|\alpha'(\psi)\>.
\]
Such a procedure gives us a way of initializing the qubit state $|0\>$ (since this is its final state regardless of the initial state) or alternatively of making a measurement of the state of the qubit, if we choose to now subject the cavity field to some type of measurement.

The above ideal evolution can be achieved if the coherent state amplitude $|\alpha|$ goes to infinity \cite{gea,attractor} - the state $|0\>$ is then known as an attractor state. The time taken, however, is proportional to $|\alpha|$ and so also diverges. We are lead therefore to wonder how well one can do if $\alpha$ is finite? This is also of interest for another reason. The laser power requirements for a large scale quantum computer capable of factorising numbers of actual interest are prodigious \cite{steane}. So a general question of interest is just how classical do the light fields need to be? This question has been considered for the case of quantum gates implemented in the Jaynes-Cummings model \cite{enk,gea}. In practise, the easiest measurements on light consist of photon counting, and we will also investigate how much information transfer is achieved when such measurements are performed. Finally, we will propose an iterative procedure for initialisation of the qubit.
%
%

\section{Jaynes-Cummings evolution}

We consider the on-resonant Jaynes-Cummings Hamiltonian in the interaction picture which takes the following form:
\begin{equation}
\label{eq-JCHamiltonian}
\hat{H}_I =  g\hslash (\hat{\sigma}_+ \otimes\hat{a} + \hat{\sigma}_-\otimes\hat{a}^\dagger),
\end{equation}
where $\hat{\sigma}_{\pm}$ are raising and lowering operators of the two-level atom and $\hat{a}$ and $\hat{a}^\dagger$ are the annihilation and creation operators of the single mode field to which it is coupled. The evolution of the state of this composite system can be determined analytically. We will focus on the reduced density matrix description of the evolution of the atom. The overall initial state is separable  and given by an arbitrary qubit state $|\psi(0)\rangle=C_g|g\rangle+C_e|e\rangle$  which is subsequently coupled to a coherent state $|\alpha\rangle =\sum_{n=0}^{\infty }\sqrt{P_{n}}|n\rangle$ with
$P_{n}(\alpha)=e^{-|\alpha|^2}\frac{\alpha^{2n}}{n!}$, where from now we take $\alpha \in \mathbb{R}^{+}$ without loss of generality. After a time $t$ the components $(x,y,z)$ of the Bloch vector $\boldsymbol{r}(t)$ describing the reduced state
\begin{eqnarray}
\label{eq-qubitreduced}
\hat{\varrho}_{\mbox{\tiny atom}}(t) &=& \mbox{Tr}_{\mbox{\tiny light}}[e^{-i \hat{H}_It/\hslash}\left(|\psi(0)\rangle\langle\psi(0)|\otimes|\alpha\rangle\langle\alpha|\right)e^{i \hat{H}_It/\hslash}]
\nonumber \\ \nonumber \\
&=& \frac{1}{2}\left(1+\boldsymbol{r}(t)\cdot\hat{\boldsymbol{\sigma}}  \right)
\end{eqnarray}
take the form (with scaled time $\tau=gt$):
\begin{widetext}
\begin{eqnarray}
\label{eq-strictblochcomponents}
x(\tau) &=&\sum_{n=0}^{\infty }P_{n}\left(2\mbox{Re} \{C_gC_e^{\ast}\}\left[\cos(\tau\sqrt{n})\cos(\tau\sqrt{n+1})
+\sqrt{\frac{n}{n+1}}\sin(\tau\sqrt{n})\sin(\tau\sqrt{n+1})\right]\right)
\nonumber\\
y(\tau) &=& \sum_{n=0}^{\infty }P_{n}\left(-2\mbox{Im}\{C_gC_e^{\ast}\}\left[\cos (\tau \sqrt{n})\cos (\tau \sqrt{n+1})
-\sqrt{\frac{n}{n+1}}\sin(\tau\sqrt{n})\sin(\tau\sqrt{n+1})\right] \right.\nonumber\\
& &\left.-2\left[|C_g|^{2}\frac{\alpha}{\sqrt{n+1}}\cos(\tau\sqrt{n})\sin(\tau\sqrt{n+1})
-|C_e|^{2}\frac{\sqrt{n}}{\alpha}\cos(\tau\sqrt{n+1})\sin(\tau\sqrt{n})\right]\right)
\nonumber\\
z(\tau) &=&\sum_{n=0}^{\infty}P_{n}\left(|C_g|^{2}\cos(2\tau\sqrt{n})-|C_e|^{2}\cos (2\tau\sqrt{n+1})
-2\mbox{Im}\{C_gC_e^{\ast}\}\frac{\sqrt{n}}{\alpha}\sin(2\tau \sqrt{n})\right).
\end{eqnarray}
\end{widetext}

\section{Information transfer to the field}

\subsection{Perfect state transfer to the field}

Let us consider the atomic reduced state for the two limits $\alpha\ll 1$ and $\alpha\gg 1$.

For $\alpha\rightarrow0$  the state of the atom asymptotically reaches a state of vacuum oscillations:
\begin{equation}\label{rho-lowalpha}
\hat{\varrho}_{\mbox{\tiny atom}}(\tau)=
\begin{pmatrix}
|C_e|^2\sin^2(\tau)+|C_g|^2
& C_gC_e^*\cos(\tau)\medskip \\
C_g^*C_e\cos(\tau)
& |C_e|^2\cos^2(\tau) \\
\end{pmatrix}
\end{equation}
At times $\tilde{\tau}_{m}=m\pi$, $m\in\mathbb{N}$, the system is in its initial state with Bloch vector $\boldsymbol{r}(\tau)=(x(0),y(0),z(0))^{T}$, but at times $\tau_k=\left(k-\frac{1}{2}\right)\pi$, $k\in\mathbb{N}$, the system evolves into the ground state $\hat{\varrho}_{\mbox{\tiny atom}}(\tau_k)=|g\rangle\langle g|$ given by $\boldsymbol{r}(\tau)=(0,0,1)^{T}$.  Since this state is independent of the the initial conditions, all information has to be transferred to the other subsystem, the light field, to conserve the information content of the total system.

For large but finite amplitudes ($\alpha\gg1$), the expressions in \eqref{eq-strictblochcomponents} can be approximated by an approach laid out in the appendix. The  components are \begin{eqnarray}
\label{eq-gauss}
x(\tau) &=& x_{\infty}\cos\left(\frac{\tau}{2\alpha}\right)e^{-\frac{\tau^{2}}{32\alpha^{4}}}
\nonumber \\ \nonumber \\
y(\tau) &=& y_{\infty}e^{-\frac{\tau {2}}{2}}-\sin\left(\frac{\tau}{2\alpha}
\right) e^{-\frac{\tau ^{2}}{32\alpha ^{4}}}
\nonumber \\ \nonumber \\
z(\tau) &=& z_{\infty}e^{-\frac{\tau ^{2}}{2}},
\end{eqnarray}
where $(x_{\infty} ,y_{\infty}, z_{\infty})$ is given by the rotation of the initial vector $(x(0), y(0), z(0))^T$ by the angle $\Omega = 2\tau\alpha$:
\begin{equation}
\begin{pmatrix}
x_{\infty } \\
y_{\infty } \\
z_{\infty }
\end{pmatrix}
=
\begin{pmatrix}
1 & 0 & 0 \\
0 & \cos (\Omega ) & -\sin (\Omega ) \\
0 & \sin (\Omega ) & \cos (\Omega )
\end{pmatrix}
\begin{pmatrix}
x(0) \\
y(0) \\
z(0)
\end{pmatrix}.
\end{equation}\\

The approach taken to arrive at the approximations \eqref{eq-gauss} holds for small $\frac{\tau}{\alpha}$.
Taking $\alpha\rightarrow\infty$ the state becomes separable when $\frac{\tau}{2\alpha} = \left(k-\frac{1}{2}\right)\pi$, $k\in\mathbb{N}$, In this limit the Bloch vector tends to  $\boldsymbol{r} = \left(0,(-1)^{k+1},0\right)$, corresponding to the attractor states of \cite{gea,attractor}.
Again, this indicates that all the information initially contained in the atomic state is perfectly transferred to the optical degrees of freedom.

\subsection{Information gain by measurement on the field}

We now investigate the quantum information transfer from the atom to the light mode occurring for arbitrary $\alpha,\tau$. For definiteness we will focus on an operationally relevant scenario - namely how much information would we expect to gain about the atomic qubit's initial state if we were to make an (idealized) photon number counting measurement on the field after some time $t$? The average information gain $I_{\mbox{\scriptsize avg}}$ is a quantity commonly used \cite{peres} to evaluate the efficiency of a measurement procedure to retrodict the quantum state - its maximum value for a qubit initially in an arbitrary state is about 0.2787. 

\begin{figure}
\begin{center}
\epsfig{file=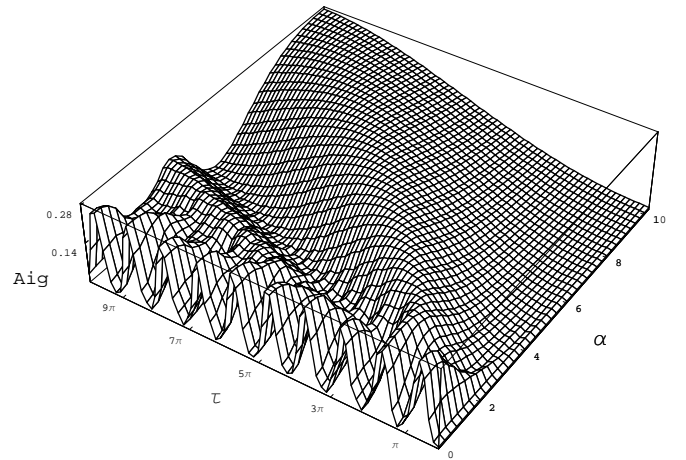, width=\columnwidth}
\caption{\label{fig-aig} Average information gain $I_{\mbox{\tiny avg}}$ as a function of time $\tau$ and amplitude $\alpha$ of the initial coherent state of light. The maximum achievable value coincides with the average information gain available in the direct measurement on the qubit.}
\end{center}
\end{figure}

To derive the average information gain about the atomic state for the case of a photon detection measurement, we do the following. The combined system of qubit and field undergoes a unitary evolution given by $e^{-i\hat{H}_I t/\hslash}$. Then, at the time $\tau$, the reduced density matrix of the qubit \eqref{eq-qubitreduced} can be also expressed as:
\begin{equation}
\hat{\varrho}_{\mbox{\tiny atom}}(t) = \sum_{n=0}^{\infty}\hat{K}_{n}(t) |\psi(0)\rangle\langle\psi(0) | \hat{K}_{n}^{\dagger}(t).
\end{equation}
The Kraus operators $\hat{K}_n(t) = \langle n|e^{-i \hat{H}_It/\hslash}|\alpha\rangle$ determining the evolution of the state of light have the form:
\begin{equation}
K_{n}=\sqrt{P_n}
\left(
\begin{array}{cc}
\cos\left(\tau\sqrt{n}\right)
& -\frac{i\sqrt{n}}{\alpha}\sin\left(\tau\sqrt{n}\right)
\\ \\-\frac{i\alpha}{\sqrt{n+1}}\sin\left(\tau\sqrt{n+1}\right)
& \cos\left(\tau \sqrt{n+1}\right)
\end{array}
\right)
\end{equation}
where $P_n=\frac{\alpha^{2n}}{n!}e^{-\alpha^{2}}$ is the probability of measuring $n$ photons in the initial state of light $|\alpha\rangle$. Each Kraus operator is actually the equivalent measurement operator describing the atomic state when a photon number measurement is performed on the field. Using the above expression we can determine the information about the initial state of the atom $|\psi(\theta, \varphi)\rangle$ gained from the photon number measurement of the optical mode that has been interacting with the atom for a time $\tau$. Here $\theta$ and $\phi$ denote the initial azimuthal and polar angle of the qubit's Bloch sphere vector. The conditional probability that for such an initial state we will detect exactly $n$ photons therefore equals:
\begin{eqnarray}
\label{eq-Conditionalonn}
P(n|\theta,\varphi)
&=& \langle\psi(\theta, \varphi)|\hat{K}^\dagger_{n} \hat{K}_{n} |\psi(\theta, \varphi)\rangle
\nonumber \\ \nonumber \\
&=& P_{n}(\alpha)\big[\cos ^{2}(\theta)f_{1}\left(n,\alpha,\tau\right)
\nonumber \\ \nonumber \\
& &+\sin^{2}\left(\theta\right)f_{2}\left( n,\alpha,\tau\right)
\nonumber \\ \nonumber \\
& &+\sin(\theta)\sin(\varphi)f_{3}\left(n,\alpha,\tau\right)\big],
\end{eqnarray}
where
\begin{eqnarray}
f_{1}\left(n,\alpha,\tau\right) &=& \cos^{2}\left(\tau\sqrt{n}\right)+\frac{\left\vert\alpha\right\vert^{2}}{n+1} \sin^{2}\left(\tau\sqrt{n+1}\right)
\nonumber \\ \nonumber \\
f_{2}\left(n,\alpha,\tau\right) &=& \cos^{2}\left(\tau\sqrt{n+1}\right)+\frac{n}{\left|\alpha\right|^{2}}\sin^{2}\left(\tau\sqrt{n}\right)
\nonumber \\ \nonumber \\
f_{3}\left(n,\alpha,\tau\right) &=& \frac{1}{2}\left(\frac{\alpha}{\sqrt{n+1}}\sin\left(2\tau\sqrt{n+1}\right)\right.
\nonumber \\ \nonumber \\
& &\left.-\frac{\sqrt{n}}{\alpha}\sin\left(2\tau\sqrt{n}\right)\right).
\nonumber \\
\end{eqnarray}
Applying Bayes' theorem
\begin{eqnarray}
\label{eq-Bayesrelation}
P(\theta,\varphi|n)&=&P(n|\theta,\varphi)\frac{P(\theta,\varphi)}{P(n)}
\end{eqnarray}
 for the conditional probability in \eqref{eq-Conditionalonn}, the average information gain in this case is given by:
\begin{align}
\label{eq-AIG}
I_{\mbox{\tiny avg}}(\tau,\alpha) &=-\int P(\theta ,\varphi )\log_2 \left( P(\theta ,\varphi )\right)\nonumber\\
&+\sum_{n=0}^{\infty }P(n)\int P(\theta ,\varphi |n)\log_2 \left( P(\theta
,\varphi |n)\right).
\end{align}
We obtain an expression that can be readily evaluated numerically. The  average information gain is a function of just $\alpha$ and $\tau$ as all other parameters have now been appropriately averaged over. Fig.~\ref{fig-aig} depicts the average information gain against $\alpha$ and $\tau$.

\begin{figure}
\begin{center}
\epsfig{file=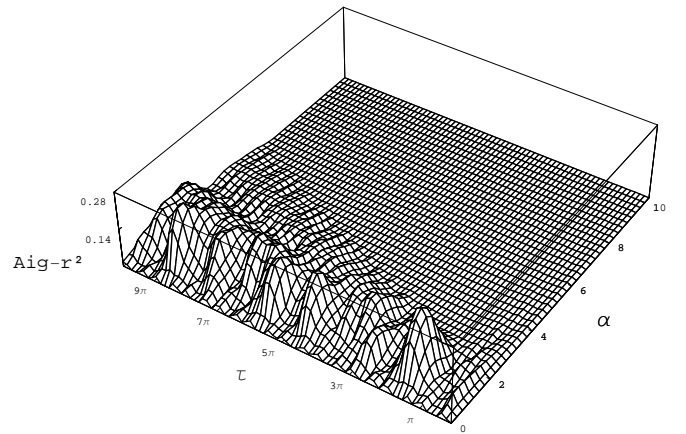, width=\columnwidth}
\caption{\label{fig-aigvsrsq} $I_{\mbox{\tiny avg}}-\mbox{\scriptsize max}(I_{\mbox{\tiny avg}})r^2$ as a function of time $\tau$ and amplitude $\alpha$ of the initial coherent state of light. It follows that both for $\alpha\gg 1$ and $\alpha\ll 1$ the average information gain can be well aproximated by the squared length of the Bloch vector.}
\end{center}
\end{figure}
\begin{figure}
\begin{center}
\epsfig{file=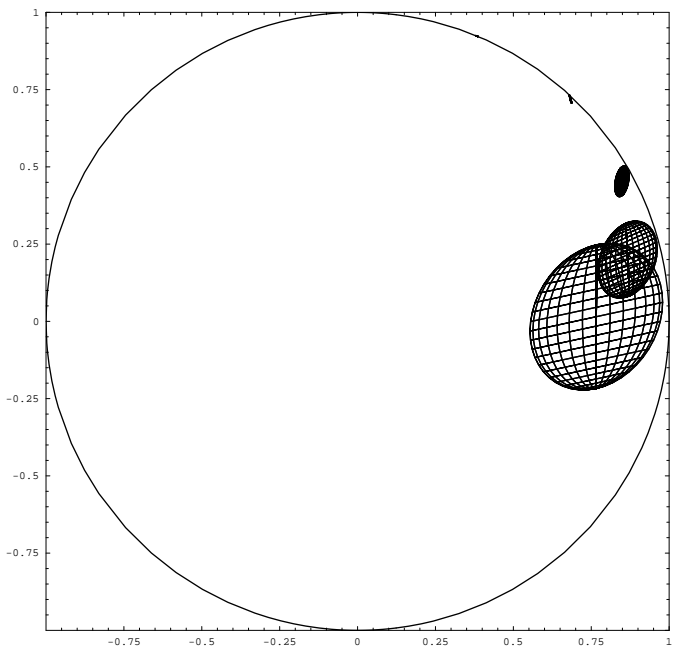, width=0.49\textwidth}
\caption{\label{fig-no4} The Bloch sphere projected onto $x=0$ is shown after an evolution of $\tau=\left(4-\frac{1}{2}\right)\pi$ for $\alpha=0.2, 0.4, 0.6, 0.8 $ and $ 1.0$\newline The contraction for $\alpha=0.2$ is so strong that it is not visible in this graph.}
\end{center}
\end{figure}

The plot confirms our conclusions based on the studies of the special cases. We observe that for certain parameters it is possible to achieve a maximal information gain equal to that of a direct measurement on the qubit. This means that the evolution, for particular times, can lead to a total transfer of information from the atomic degree of freedom to the optical one.

The first attractor state is a notable feature of the plot. It is described by the collection of points $(\tau=\pi\alpha,\alpha)$. The second attractor state can be made out as the smoothed ridge along $(\tau=3\pi\alpha,\alpha)$. The value of the average information gain increases monotonically along these lines as $\alpha$ increases, which is to be expected given the from of equation \eqref{eq-gauss}.

Another feature we observe are the vacuum oscillations. As formula \eqref{rho-lowalpha} suggests, maxima of the average information gain occur periodically for \break$(\tau=\left(k-\frac{1}{2}\right)\pi,0)$ and $k\in\mathbb{N}$. An interesting feature is the extent of these local maxima for constant time at increasing values of $\alpha$. In the parameter plane these points are $(\tau=\left(k-\frac{1}{2}\right)\pi,\alpha)$ and $k\in\mathbb{N}$.

These findings suggest that there are states parameterised by $(\tau,\alpha)$ along these local maxima of $I_{\mbox{\tiny avg}}$ which are almost separable, i.e. lie in the periphery of the Bloch sphere.
The question arising is: which almost separable states do they represent, since in the low alpha limit they become the ground state \eqref{rho-lowalpha} and in the high alpha limit they become the attractor states \eqref{eq-gauss}?
This question is addressed in the next section\ref{blochball}.

To conclude the discussion of the average information gain we would like to point out a useful correspondence between the radius of the average state, here given implicitly:\begin{equation}\label{averagerho}
\frac{1}{4\pi}\int_{\Omega}\hat{\varrho}_{\mbox{\tiny atom}}(\tau,\theta,\phi)d\Omega= \frac{1}{2}\left(1+\langle\boldsymbol{r}(\tau)\rangle\cdot\hat{\boldsymbol{\sigma}}  \right)
\end{equation}
 and the average information gain $I_{\mbox{\tiny avg}}$.

Fig.~\ref{fig-aigvsrsq}
shows this correspondence: $I_{\mbox{\tiny avg}}$ can
be well approximated by \eqref{averagerho}. The difference $I_{\mbox{\tiny avg}}-\mbox{\scriptsize max}(I_{\mbox{\tiny avg}})\langle\boldsymbol{r}(\tau)\rangle^2$ is close to zero both for very small and very large amplitudes $\alpha$. Only for intermediate values $\alpha\approx 1$ the difference is essentially non-zero. However, even in this case the oscillations of the length $r^2$ uniquely correspond to the oscillations of the average information gain $I_{\mbox{\tiny avg}}$ in the mode of light. This indicates that instead of carrying out numerical analysis one can apply \eqref{averagerho} to obtain nearly identical results.

\section{Qubit Initialisation}\label{blochball}

In this section we perform a detailed analysis of how one may unitarily (completely coherently) ``pump'' an atom into a fixed desired state, regardless of its initial state. This procedure of ``qubit initialization'' is important for quantum computing applications, and is normally done using incoherent processes and ancillary levels. We investigate those  states that we have identified in the previous section as almost separable by means of the average information gain. A qubit that can be predicted with a high probability to be in a particular state irrespectively of its initial state is effectively initialised. In particular we consider iterative procedures for controlling the initialization to a wide range of desired states. 

At constant $\tau=\left(k-\frac{1}{2}\right)\pi$, $k\in\mathbb{N}$ and along $\alpha \in (0,1]$ the Bloch sphere contracts down to regions close to the periphery in a clockwise or anticlockwise fashion depending on $k$.
For $k=4$, i.e. $\tau=\frac{7}{2}\pi$, the shrunken Bloch spheres for values of $\alpha\leq1$ are depicted in Fig.~\ref{fig-no4}. It can be seen that
the spheres propagate along the periphery from $\theta=0$ to $\theta=\pi/2$ as $\alpha$ goes from 0 to 1. The proximity of the spheres to the pure states of radius 1 can be ascribed to the high value of the average information gain that was found earlier.

How pure can we get? It would be a desired effect in an initialisation procedure for the degree of purity of the qubit to be enhanced. We propose to successively evolve the system with a specially chosen set of parameters. An interesting case is the iterative application of the radiation field with the same set of parameters $\alpha$ and $\tau$ respectively:
\begin{equation}\label{iterative}
\hat{\varrho}_{\mbox{\tiny{\it{N}}}}= \sum_{n_{1},...,n_{N}=0}^{\infty}\left[\hat{K}_{n_{N}}\cdot\cdot\cdot\hat{K}_{n_{1}}\right]\hat{\varrho}_{\mbox{\tiny{0}}}\left[\hat{K}_{n_{1}}^{\dagger}\cdot\cdot\cdot\hat{K}_{n_{N}}^{\dagger}\right]
\end{equation}
where $\hat{K}_{n_{i}}=\hat{K}_{n_{i}}(\tau,\alpha)$ are the Kraus operators which only depend on $(\tau,\alpha)$ for each field $i\in\{1,...,N\}$ and $\hat{\varrho}_{\mbox{\tiny{0}}}=\hat{\varrho}_{\mbox{\tiny{atom}}}(0) $ is the initial state of the atom.

\begin{figure}
\begin{center}
\epsfig{file=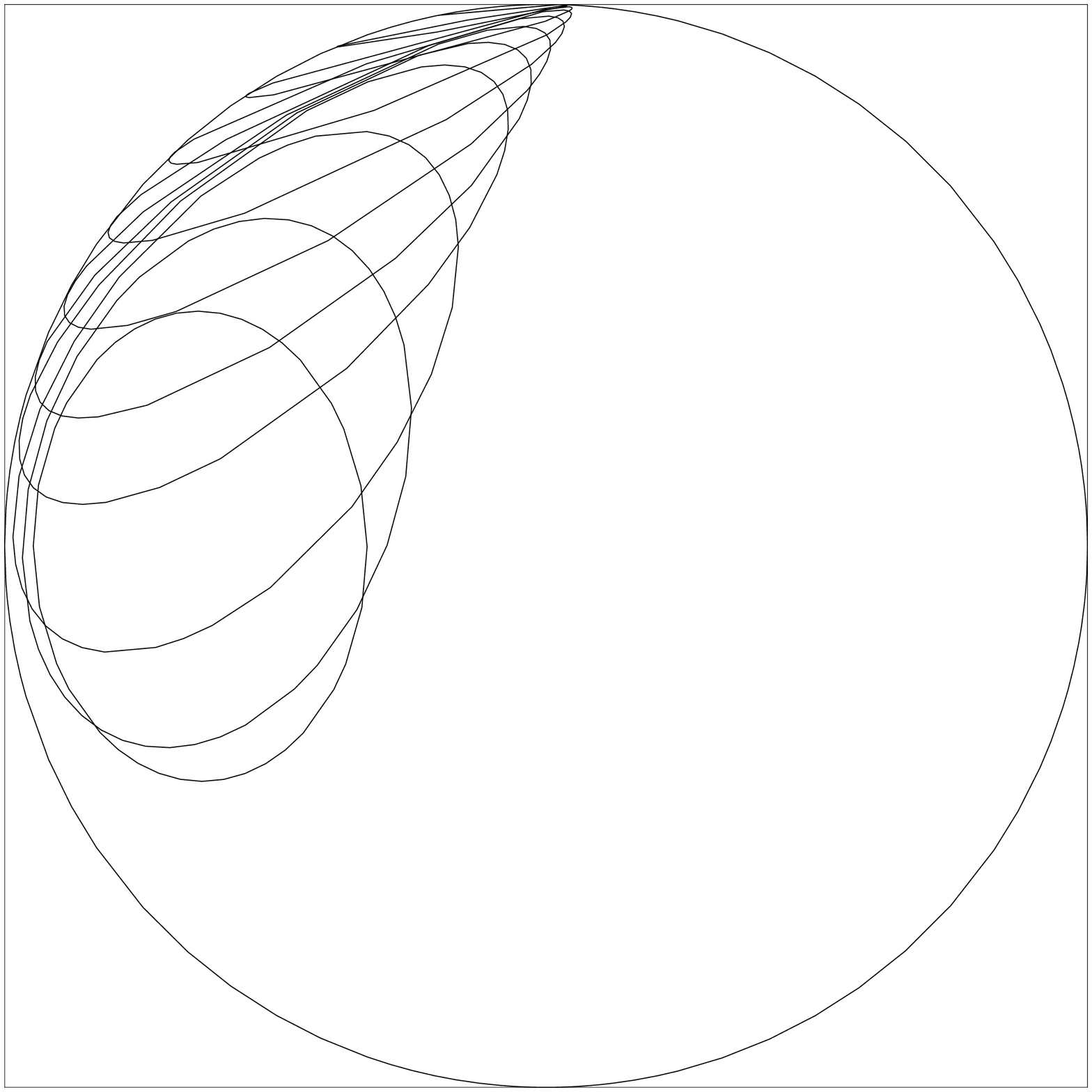, width=2.8cm}
\epsfig{file=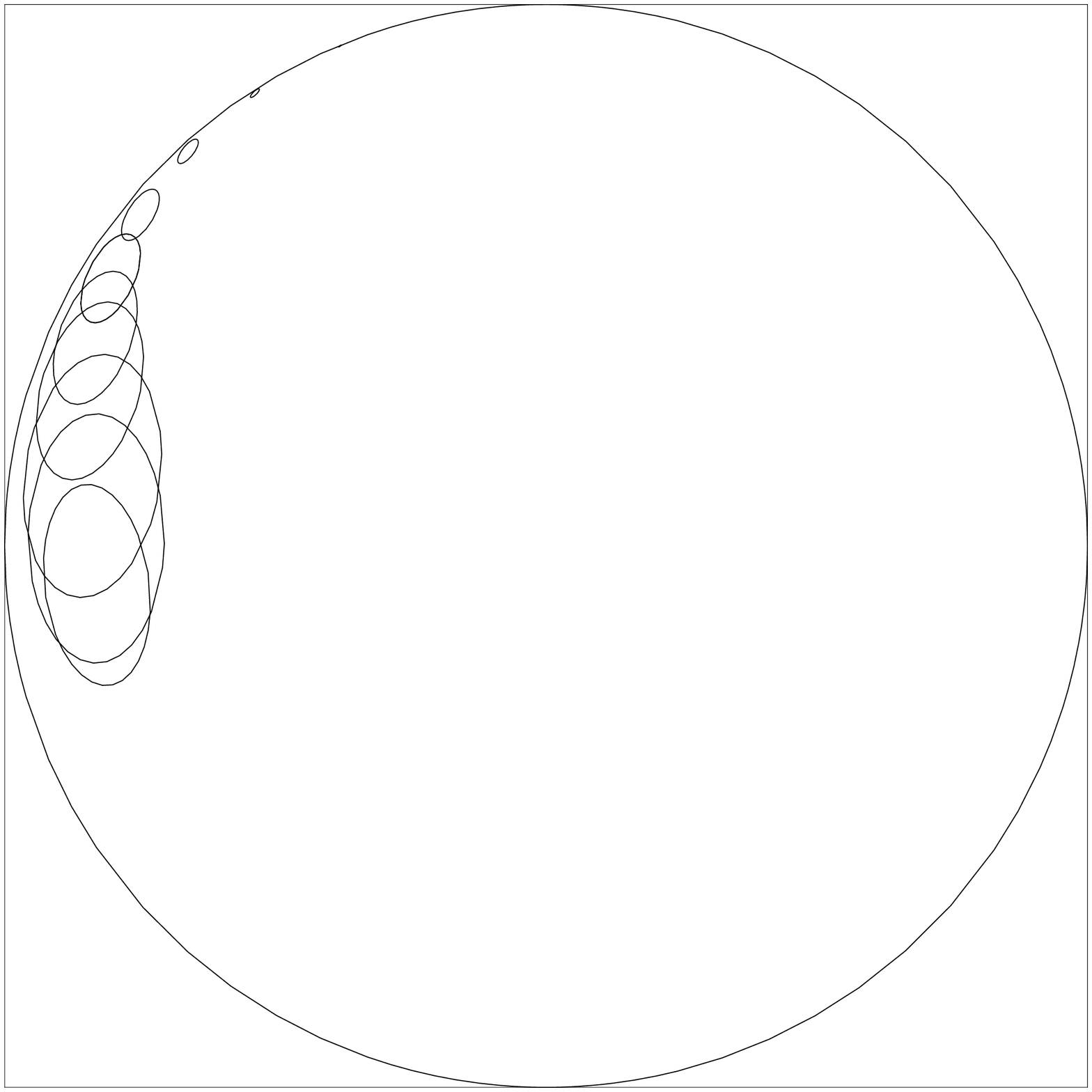, width=2.8cm}
\epsfig{file=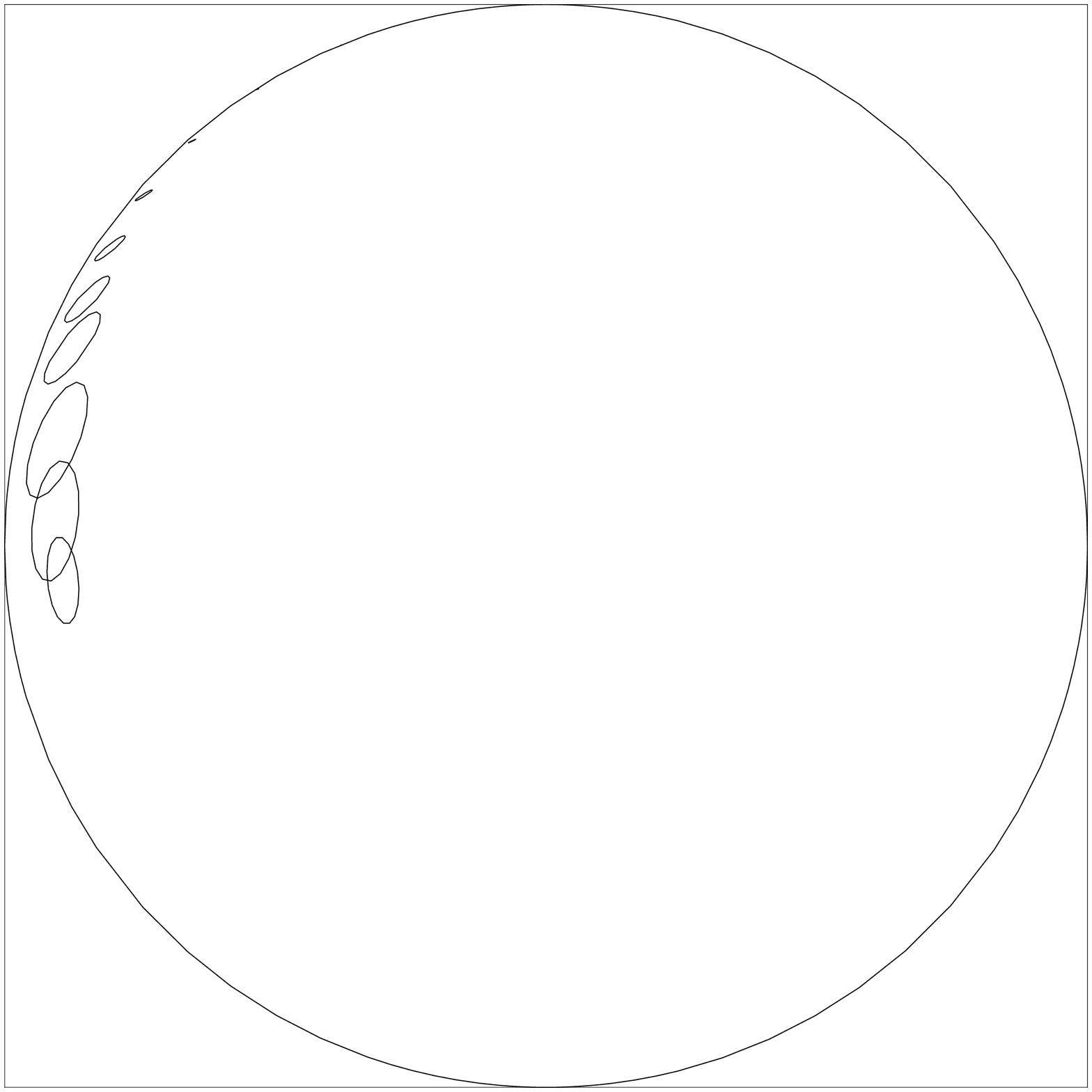, width=2.8cm}
\epsfig{file=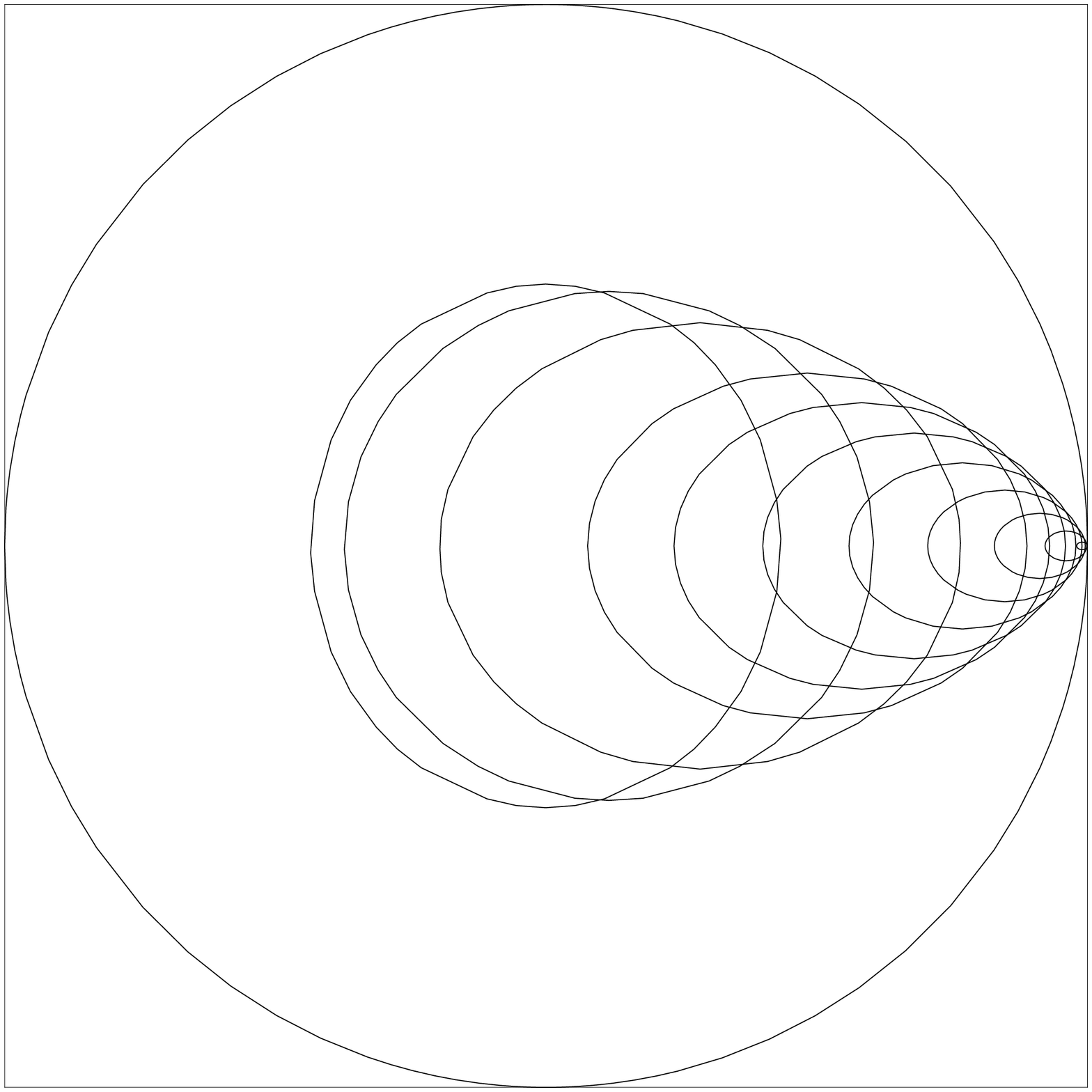, width=2.8cm}
\epsfig{file=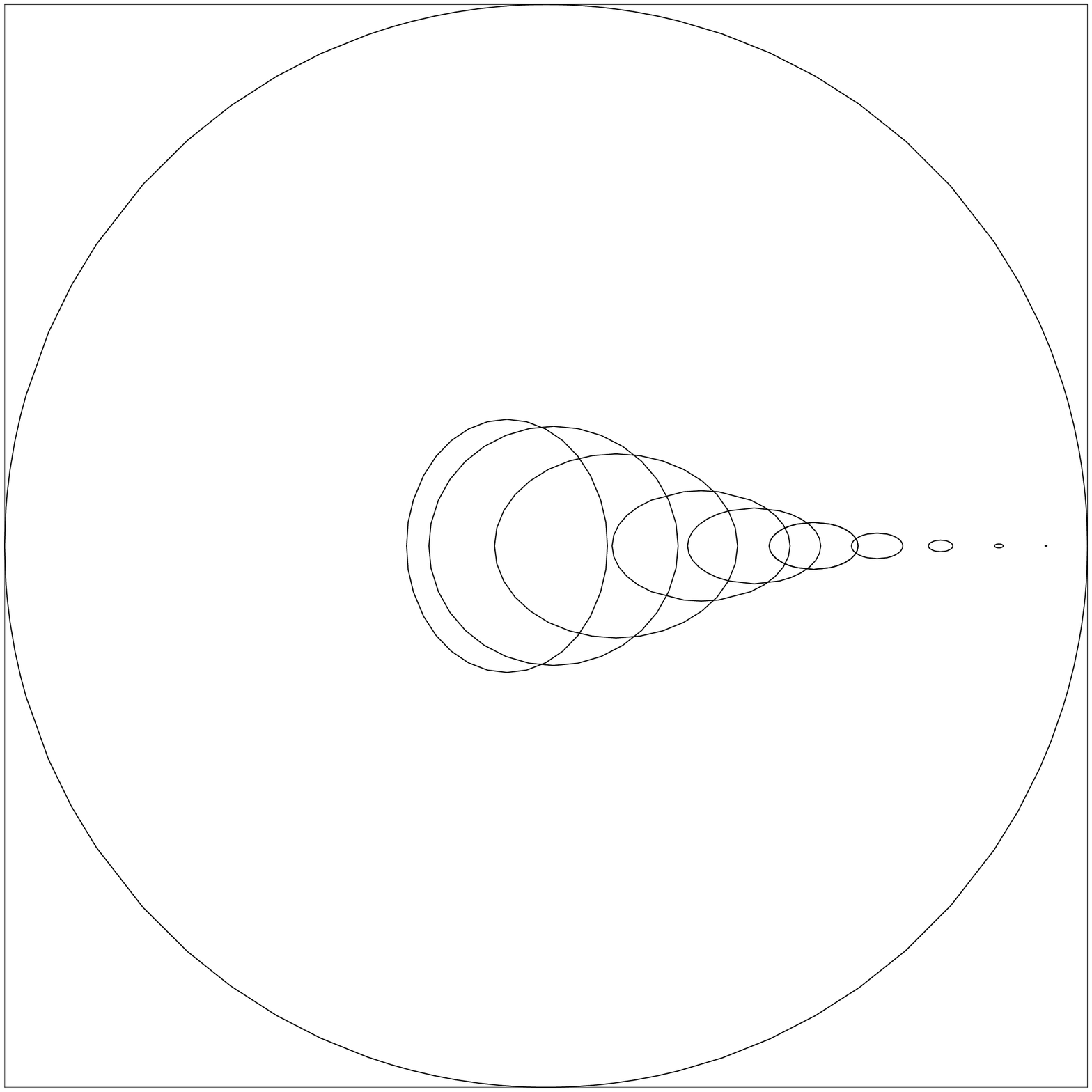, width=2.8cm}
\epsfig{file=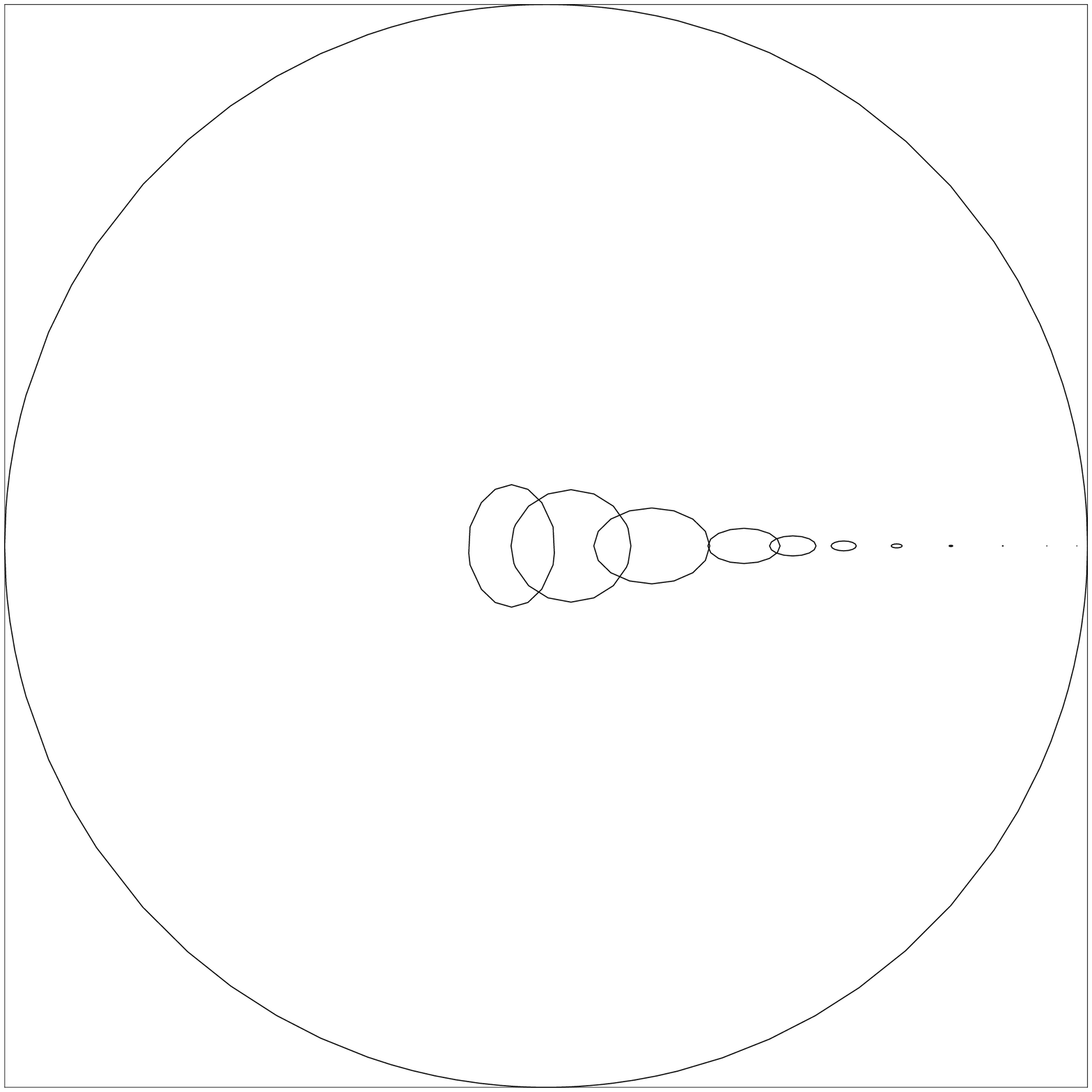, width=2.8cm}
\epsfig{file=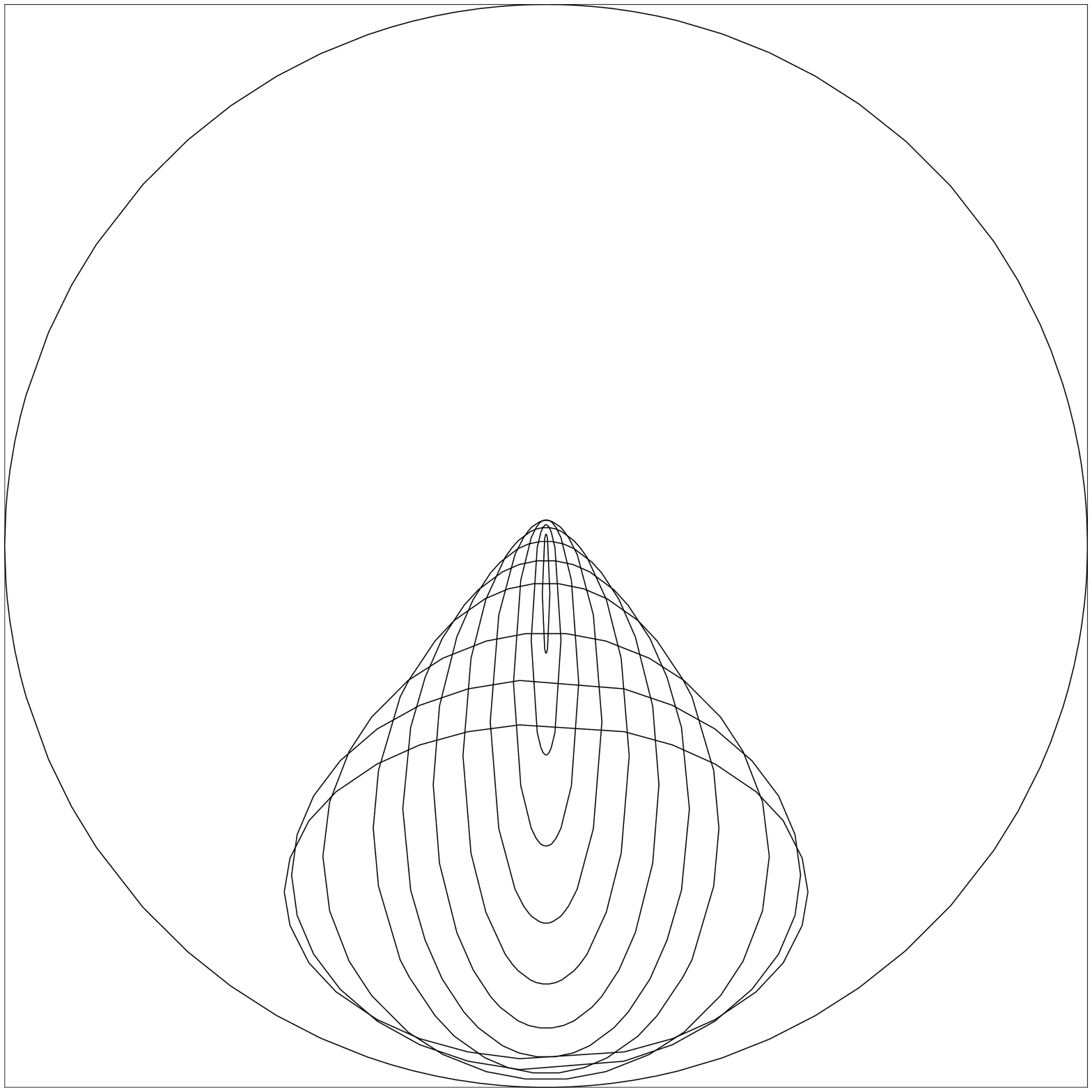, width=2.8cm}
\epsfig{file=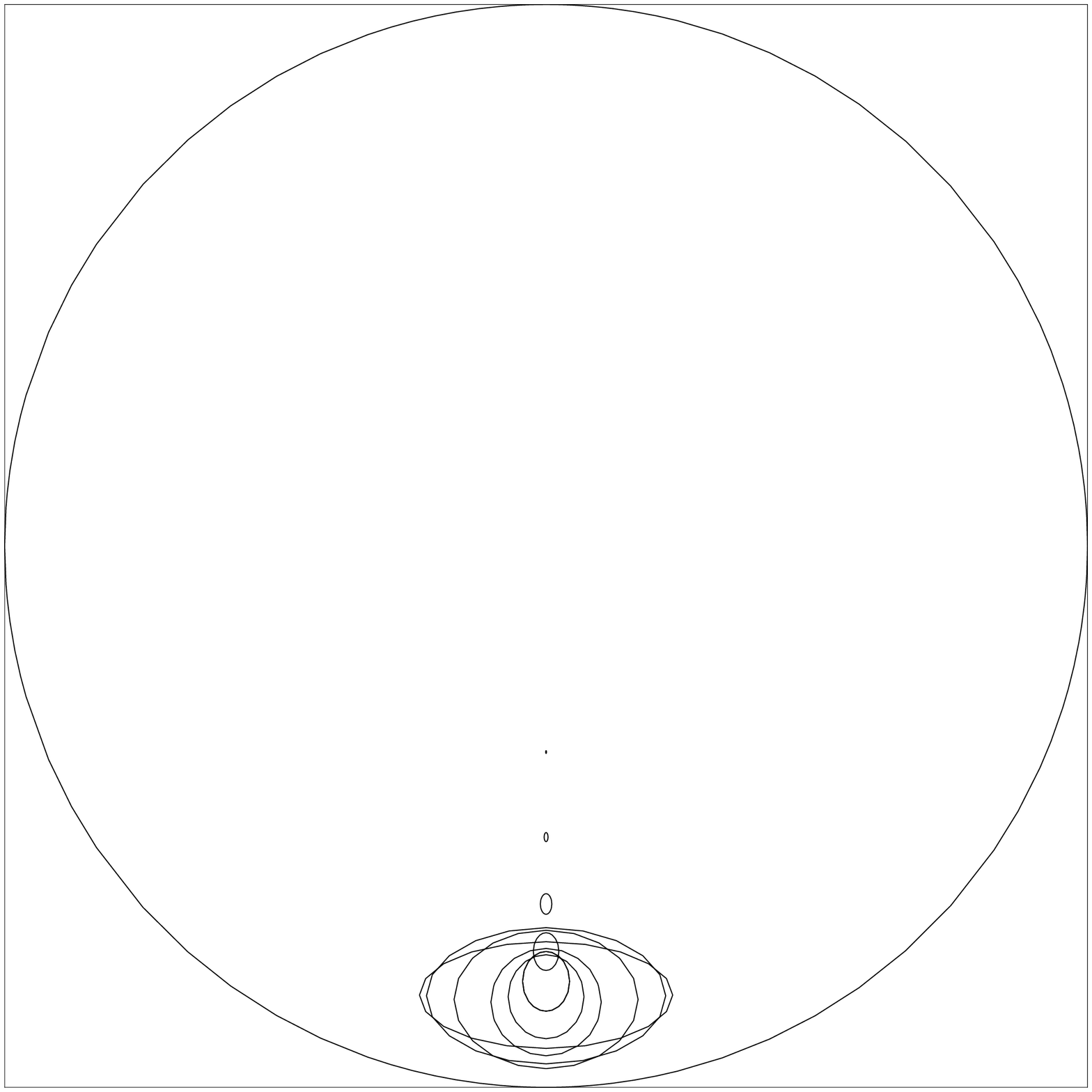, width=2.8cm}
\epsfig{file=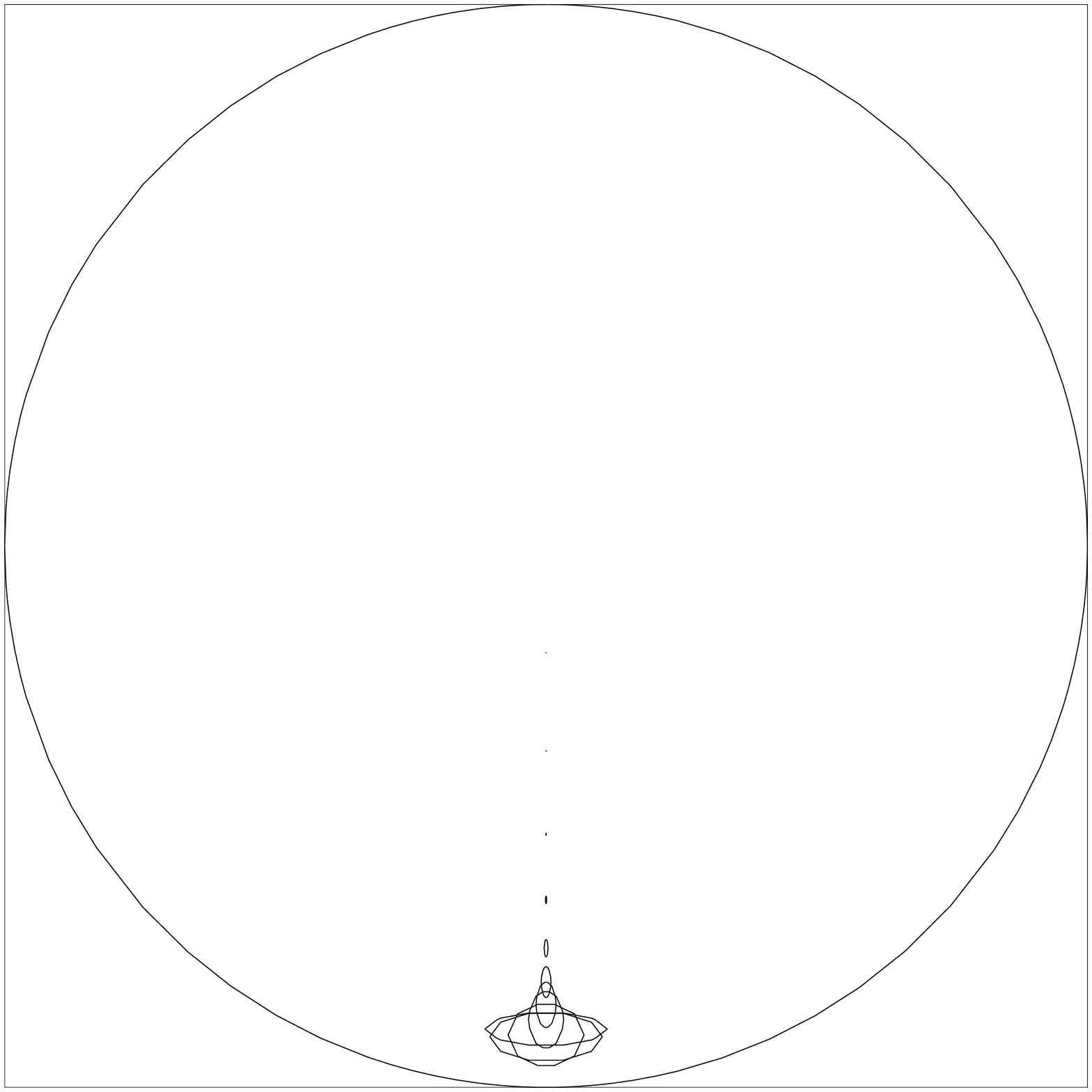, width=2.8cm}
\caption{\label{fig-no3}  Initialisation procedure for $\tau=\frac{5}{2}\pi$: Columns correspond to iteration steps (from left to right: one, two and three applications of a field). Every row corresponds to a different 2D projection of the Bloch sphere (x, y and z respecively).}
\end{center}
\end{figure}
This procedure is shown in Fig.~\ref{fig-no3} for a radiation time of $\tau=\frac{5}{2}\pi$, corresponding to $k=3$, and a set of coherent amplitudes $\alpha\in\{0.2,0.4,0.6,0.8,1.0\} $. Three iterations were calculated numerically where each column represents one iteration and each row a different point of view on the Bloch sphere.

The degree of purity is enhanced with each iteration as can be seen clearly  from Fig.~\ref{fig-no3}. Similar to the observation done for the case of $\tau=\frac{7}{2}\pi$, we see that the attained states wander from  $\theta=0$ to $\theta=-\pi/2$ as $\alpha$ goes from 0 to 1. Thus all states $|\psi(\theta)\rangle=\cos(\theta/2)|0\rangle-i\sin(\theta/2)|1\rangle$ with $0\leq\theta\leq\pi/2$ are attained approximately.

The choice of the meridian can also be adjusted by the choice of the phase of the complex amplitude \break$\alpha=\alpha_{r}e^{i\phi}$. The axis of rotation thus changes from $\hat{\boldsymbol{x}}$ to any axis in the equatorial plane: $\hat{\boldsymbol{x}}\cos{\phi}+\hat{\boldsymbol{y}}\sin{\phi}$.

By this means an arbitrary state on the ground state hemisphere could be achieved as an initial state of a quantum computation. This is a practical example of the generalized SWAP gate that transmits information between subsystems of discrete and continuous degrees of freedom.

\section{Conclusions}

We have investigated the transfer of quantum information between a two-level atom and a single mode field for finite field strengths. We focussed on the purity of the state of an atom in a coherent field in the Jaynes-Cummings-model by means of the average information gain for finite values of $\alpha$ in the low and high $\alpha$ limits, providing useful analytical approximations where necessary. This has led us to find states in the ground state hemisphere to  which an atom can be initialised  by choosing the correct parameters for interaction time $\tau$ and field strength $\alpha$.  Based on these findings we have proposed a novel method of ``attractor state'' style initialization with a broader range of possible initialisation states, involving the iterative application of the radiation field.

\section*{Appendix}

In the following we will give a derivation of the approximation \eqref{eq-gauss}. For large values of $\alpha $, an approach according to \cite{intro_qo} can
be generalized in order to simplify
the infinite sums to finite expressions. The basis of this method is a
linear expansion of the frequencies $\omega (n)$ around the mean photon
number $\alpha ^{2}$:
\begin{eqnarray}
\omega (n) &\approx &\omega (\alpha ^{2})+\frac{d\omega }{dn}|_{\alpha
^{2}}(n-\alpha ^{2})  \notag \\
&\equiv &\omega _{0}+\beta (n-\alpha ^{2}).  \label{lin_approx}
\end{eqnarray}

In the case of a cosine function, this gives
\begin{widetext}
\begin{eqnarray}
\label{eq-gaussapprox}
\sum_{n=0}^\infty P_n\cos(\omega \tau) &=&e^{-\alpha^2}\sum_{n=0}^\infty
\frac{\alpha^{2n}}{n!}\left(\frac{e^{i\omega \tau}+e^{-i\omega \tau}}{2}%
\right)  \notag \\
&\approx&\frac{e^{-\alpha^2}}{2}\sum_{n=0}^\infty \frac{\alpha^{2n}}{n!}
\left(e^{i(\omega_0-\beta\alpha^2)\tau}e^{i\beta n\tau}
+e^{-i(\omega_0-\beta\alpha^2)\tau}e^{-i\beta n\tau}\right)  \notag \\
&=&\frac{e^{-\alpha^2}}{2} \left(e^{i(\omega_0-\beta\alpha^2)\tau}\exp\left[%
\alpha^2e^{i\beta \tau}\right] +e^{-i(\omega_0-\beta\alpha^2)\tau}\exp\left[%
\alpha^2e^{-i\beta \tau}\right]\right)\notag\\
&\approx&\frac{e^{-\alpha^2}}{2}
\left(e^{i(\omega_0-\beta\alpha^2)\tau}e^{\alpha^2+i\beta\alpha^2\tau}
+e^{-i(\omega_0-\beta\alpha^2)\tau}e^{\alpha^2-i\beta\alpha^2\tau}\right)e^{-%
\frac{1}{2}\beta^2\alpha^2\tau^2}  \notag \\
&=&\cos(\omega_0\tau)e^{-\frac{1}{2}\beta^2\alpha^2\tau^2},
\end{eqnarray}
\end{widetext}
where in the second last line an additional expansion $e^{\pm i\beta \tau}\approx1\pm i\beta \tau-%
\frac{\beta^2\tau^2}{2}$ was used, which is valid for small\footnote{The model will still be valid for large time scales as long as $|\beta|$ is
small, i.e. for times $\tau\ll\frac{1}{|\beta|}$} values of $\beta\tau$.

Using the approximation $\sqrt{P_{n\pm 1}}\approx \sqrt{P_{n}}$ and therefore $\frac{%
\alpha }{\sqrt{n+1}}\approx \frac{\sqrt{n}}{\alpha }\approx \sqrt{\frac{n}{%
n+1}}\approx 1$ in equations \eqref{eq-strictblochcomponents} isolates
the interference effect of the different frequencies $\sqrt{n}$ and $\sqrt{%
n+1}$ to give Bloch components
\begin{align}
\label{eq-gaussapprox2}
x&\approx\sum_{n=0}^\infty P_n \Bigl[x_0\cos\left(\tau\omega_-\right)%
\Bigr]\notag \\
y&\approx\sum_{n=0}^\infty P_n \Bigl[y_0\cos\left(\tau\omega_+\right)-
z_0\sin\left(\tau\omega_+\right)-\sin\left(\tau\omega_-\right)\Bigr]
\notag \\
z&\approx\sum_{n=0}^\infty P_n \Bigl[y_0\sin(\tau\omega_n)+z_0\cos(\tau%
\omega_n)\Bigr],
\end{align}
where the frequencies are $\omega_n\equiv2\sqrt{n}$ and $\omega_\pm\equiv
\sqrt{n+1}\pm\sqrt{n}$ and the initially pure state is given by $x_0=2\mathbf{R}\mathbf{e}\{C_gC_e^*\}$, $y_0=-2\mathbf{I}\mathbf{m}\{C_gC_e^*\}$ and $z_0=|C_g|^2-|C_e|^2$.

Applying \eqref{eq-gaussapprox} to these three equations yields different parameters $\beta$ for the Gaussian envelopes. Explicitly, they are given by
\begin{align}
\beta_n&\equiv\frac{d\omega_n}{dn}\bigg|_{\alpha^2}=\frac{1}{\alpha}~.\\
\intertext{For large $\alpha$, the following approximations are useful:}
\omega_{n_+}&\approx\omega_{n_0}=2\alpha  \label{omega_fast} \\
\omega_{n_-}&\approx\frac{1}{2\alpha}  \label{omega_slow} \\
\beta_+&\approx\frac{1}{\alpha}=\beta_n \\
\text{and }\,\,\,\,\,\,\beta_-&\approx-\frac{1}{4\alpha^3}.
\end{align}
where $\omega_{n_0}$ and $\omega_{n_\pm}$ correspond to the frequency
$\omega_0$ of the oscillation in \eqref{eq-gaussapprox}. The Bloch components can therefore be written as
\begin{align}
\label{eq-gaussapprox3}
x &\approx x_{0}\cos \left( \frac{\tau }{2\alpha }\right)
e^{-\frac{1}{2}\beta _{-}^{2}\tau ^{2}\alpha ^{2}}\notag
\\
y &\approx \Bigl[y_{0}\cos \left( 2\alpha \tau \right) -z_{0}\sin \left(
2\alpha \tau \right) \Bigr]e^{-\frac{1}{2}\beta _{+}^{2}\tau ^{2}\alpha
^{2}}\notag
\\
&\hspace{14pt}-\sin \left( \frac{\tau }{2\alpha }\right) e^{-\frac{1}{2}\beta
_{-}^{2}\tau ^{2}\alpha ^{2}} \notag
\\
z &\approx \Bigl[y_{0}\sin (2\alpha \tau )+z_{0}\cos (2\alpha \tau )\Bigr]%
e^{-\frac{1}{2}\beta _{n}^{2}\tau ^{2}\alpha ^{2}},
\end{align}
which is equivalent to equations \eqref{eq-strictblochcomponents}.

\end{document}